\documentclass[10pt, conference, letterpaper]{IEEEtran}

\usepackage{url}
\usepackage{graphicx,epsfig}% Include figure files
\usepackage{subfig}
\usepackage{graphics,dcolumn,bm,epic, eepic,fleqn,float}
\usepackage{amssymb,amsmath,multirow,rotate,color}
\usepackage{epstopdf}
\usepackage{soul}
\usepackage[draft]{hyperref}
\usepackage{color}
\usepackage[]{algorithm2e}
\usepackage{booktabs}
\usepackage[T1]{fontenc}

\newcommand\fref[1]{Fig.~\ref{#1}}
\newcommand\tref[1]{Table~\ref{#1}}

\let\oldenumerate\enumerate
\renewcommand{\enumerate}{
  \oldenumerate
  \setlength{\itemsep}{1pt}
  \setlength{\parskip}{0pt}
  \setlength{\parsep}{0pt}
}

\let\olditemize\itemize
\renewcommand{\itemize}{
  \olditemize
  \setlength{\itemsep}{1pt}
  \setlength{\parskip}{0pt}
  \setlength{\parsep}{0pt}
}

\begin{document}

\title{ISP-friendly Peer-assisted On-demand Streaming of Long Duration Content in BBC iPlayer}

\author{\IEEEauthorblockN{Dmytro Karamshuk\IEEEauthorrefmark{1},
Nishanth Sastry\IEEEauthorrefmark{1},
Andrew Secker\IEEEauthorrefmark{2} and 
Jigna Chandaria\IEEEauthorrefmark{2}}
\IEEEauthorblockA{\IEEEauthorrefmark{1}Department of Informatics, King's College London, UK. 
Email: {firstname.lastname}@kcl.ac.uk}
\IEEEauthorblockA{\IEEEauthorrefmark{2}BBC R\&D, London, UK. Email: {firstname.lastname}@rd.bbc.co.uk}
}

\maketitle

\begin{abstract}

In search of scalable solutions, CDNs  are exploring P2P support. However, the  benefits of peer assistance can be limited by various obstacle factors such as \emph{ISP friendliness}---requiring peers to be within the same ISP, \emph{bitrate stratification}---the need to match peers with others needing similar bitrate, and \emph{partial participation}---some peers choosing not to redistribute content. 

This work relates potential gains from peer assistance to the average number of users in a swarm, its \emph{capacity}, and empirically studies the effects of these obstacle factors at scale, using a month-long trace of over 2 million users in London accessing BBC shows online. Results indicate that even when P2P swarms are localised within ISPs, up to $88$\% of traffic can be saved. Surprisingly, bitrate stratification results in 2 large sub-swarms and does not significantly affect savings. However, partial participation, and the need for a minimum swarm size do affect gains. We investigate improvements to gain from increasing content availability through two well-studied techniques: \emph{content bundling}--combining multiple items to increase availability, and \emph{historical caching} of previously watched items. Bundling proves ineffective as increased server traffic from larger bundles outweighs benefits of availability, but simple caching can considerably boost traffic gains from peer assistance. 

\end{abstract}
\section{Introduction}

In recent years, the rise of multimedia streaming and Content Delivery Networks (CDNs) has led to a decrease in the popularity of peer-to-peer content downloads~\cite{labovitz2010internet}. Curiously, there has been a simultaneous surge of interest among CDN operators in using hybrid peer-assisted approaches to offload some of their server traffic. Early feasibility studies using MSN video traces revealed that substantial savings could be obtained, \emph{for the two most popular videos in the trace}~\cite{huang2007can}. Recent large-scale measurements suggest that such approaches might be extremely effective across entire content corpus, with a reported 70\% of server traffic offloaded worldwide in Akamai's NetSession~\cite{zhao2013peer}, and over 87\% savings for a USA-wide Video-on-Demand (VoD) workload from Conviva~\cite{balachandran2013analyzing}. Contrariwise, there have been several large-scale deployments of peer-to-peer (P2P) streaming systems such as GridCast~\cite{cheng2008gridcast} and UUSee~\cite{liu2010uusee}, which report the need for various degrees of server assistance. 

Thus, there appears to be a clear consensus that P2P support can greatly decrease the cost of content delivery. However, there are still several obstacles in the details: For instance, \emph{ISP friendliness} has been a potential point of concern as peers from  different ISPs exchanging content can increase of each ISP's transit traffic costs~\cite{karagiannis2005should}. Further, especially in the case of large national ISPs, there may be a need to match peers within the same region or city. Although several locality-aware approaches have been proposed~\cite{karagiannis2005should,xie2008p4p,choffnes2008taming,le2011pushing,cuevas2013bittorrent}, measurements have shown that finding local peers may not always be easy~\cite{piatek2009pitfalls}. In NetSession, only $\approx$18\% of P2P traffic remains within the same Autonomous System (AS)~\cite{zhao2013peer}; and in the Conviva traces, server traffic savings drop from 87\% to 13\% if swarms are restricted to peers within the same ISP and city~\cite{balachandran2013analyzing}. Further, the swarming capacity of peers may be limited because of asymmetry in upload/download bandwidths, as well as a general reluctance of some users to upload content, or change settings to allow background uploads (only 31\% of NetSession clients have upload enabled~\cite{zhao2013peer}). Thus, there may only be a \emph{partial participation} from peers, whereby the collective traffic contributed  to the swarm is only a fraction of the consumption levels. Applications such as multimedia streaming face additional difficulties in maintaining stable swarms due to \emph{bitrate stratification}, because peers may need different bitrates at different times depending on current network conditions. 

We approach these issues in the context of \emph{on-demand} streaming of \emph{long duration} multimedia content such as TV shows and movies. The sheer size of long duration content makes this one of the largest class of applications on the Internet today. For example, in the USA, Netflix makes up a reported 32.7\% of peak time traffic~\cite{sandvine}. Thus, confirming the gains from P2P approaches in this setting could go a long way towards making hybrid CDNs more mainstream in today's content delivery architectures. 

Intuitively, the success or not of peer assistance depends on the tension between two factors. On the one hand, on-demand streaming has been a difficult case for P2P approaches because of  potential asynchronicity in peer arrival times. On the other hand, we make the observation that in today's streaming model, users remain online as long as they are watching the content. If this ``online while you watch'' model is preserved in the peer-assisted approach, the \emph{availability} of content, a key factor in the efficiency of P2P swarming~\cite{menasche2013content,kaune2010unraveling}, improves dramatically due to the long duration of content, and could potentially offset differences in peer arrival times.

\begin{table}
\small{
\centering
\begin{tabular}{c|c}
Parameter & Value \\
\hline
\hline
Number of Users & 2.2M \\
Number of IP addresses & 1.3M \\
Number of Sessions & 15.9M \\
\hline
\end{tabular}
\caption{Accesses to BBC iPlayer in London, Sep 2013.}\vspace{-6mm}
\label{tbl:dataset}}
\end{table}

We empirically examine whether, at scale, the balance tilts in favour of peer assistance, using a month-long trace of almost 16 Million sessions of BBC iPlayer. iPlayer is a ``catch-up'' streaming service which allows on-demand streaming of TV and radio shows recently broadcast by the British Broadcasting Corporation (BBC) in the United Kingdom (UK). Started in 2008, by 2012, it had been used been used by an estimated 44\% of UK households~\cite{ofcom}, and was the most popular long-duration content streaming application in the UK, second only to YouTube amongst all streaming sources~\cite{sandvine}. Our trace contains 1.2 Million IPs from more than 2 Million users located in one city, London, allowing us to capture well the locality issues discussed above. %The IPs we observe correspond to $\approx$36\% of London's 3.3 Million households, and the users number $\approx$ 25\% of London's population\footnote{\url{http://data.london.gov.uk/documents/focus-on-london-2011-housing.pdf}}. 
Dataset parameters are summarised in \tref{tbl:dataset}. 
%http://www.kgbanswers.co.uk/how-many-households-are-there-in-london/3471418 (there are 5,183,970 households in London)
 %There are 26.4 million households. 
% Of course, each person has more than one IP they are associated with.
%http://www.ons.gov.uk/ons/rel/family-demography/families-and-households/2013/stb-families.html

Although iPlayer is currently an over-the-top streaming service using traditional CDNs, we use trace-driven simulations to explore the potential advantage of a hybrid P2P CDN in comparison with a streaming-only CDN, in terms of its \emph{traffic gain}, the fraction of the users' traffic that is offloaded from the server via peer assistance. We first investigate whether and to what extent traffic gain is affected by the above ``obstacle factors''---ISP friendliness, partial participation, bitrate stratification and asynchronicity in peer accesses, given the scale of iPlayer and the consequent large swarming capacity, and the higher availability of the ``online while you watch'' model. Next, we explore the relationship between increasing content availability and improvements in traffic gain. Specifically, we compare proactive approaches such as bundling~\cite{menasche2013content} and reactive pull-based approaches such as caching, both of which have been widely used to improve availability in peer-to-peer systems. Our findings may be summarised as follows:
%\begin{itemize}\itemsep1pt \parskip0pt \parsep0pt \labelindent0pt
%\item  
A few highly popular items (e.g., items with $>100K$ sessions) can obtain gains of nearly 99\% in the best case, and are hardly affected by the obstacle factors. Less popular items are affected to varying extents; unpopular items with about $1K$ sessions may see a gain of less than 20\% even in one of the Top-5 ISPs by size. The ``online while you watch'' model is critical: Traffic gains can more than halve even for popular items, in the presence of high bandwidth peers who can quickly download an item and depart from the swarm.
%  \item 
Among the obstacles considered, partial participation affects gain the most since we assume it uniformly decreases  swarm size. Others obstacles are less critical because they create one or two large sub-groups from the original swarm; and large groups have sufficient peer upload capacity, allowing for effective content interchange. For example, dividing peers based on their ISP still creates several large swarms, which together account for a large proportion of users and sessions. A surprising  finding for us was that bitrate changes are relatively uncommon, and two bitrates account for $74\%$ of sessions; thus rendering bitrate stratification ineffective as an obstacle. 
%    \item 

As one corollary of the large sub-groups, gains across the  content corpus remain relatively high; up to 88\% of traffic can be saved on average despite obstacles. The high system-wide gains are also a result of skewed popularity---the top 5\% of items account for 80\% of traffic, allowing large stable swarms of over 50 or 100 peers at a time even after obstacle factors.
%  \item 
Straightforward caching of a handful ($\le 10$) of recently watched objects is highly effective and can improve swarming capacity by 10x on average across the content corpus, translating to an up to 23\% gain improvement.

%   [30/03/2014 00:24:53] Nishanth Sastry: I see the 15% for the blue line in fig 6a
% [30/03/2014 00:24:57] Nishanth Sastry: which is 30%
% [30/03/2014 00:25:25] Dmytro Karamshuk: 86%
% [30/03/2014 00:25:31] Dmytro Karamshuk: after
% [30/03/2014 00:25:44] Dmytro Karamshuk: 67% before
% [30/03/2014 00:25:52] Nishanth Sastry: ah you are doing (86-67)/67
% [30/03/2014 00:25:54] Nishanth Sastry: OK
% [30/03/2014 00:26:03] Nishanth Sastry: I was just saying 67 + 15%
% [30/03/2014 00:26:04] Dmytro Karamshuk: if you say "by x%"
% [30/03/2014 00:26:23] Dmytro Karamshuk: I think it's better to say  (86-67)/67
% [30/03/2014 00:26:28] Dmytro Karamshuk: I alway do this mistake
%  \item 
Surprisingly, bundling, which has been shown to be highly effective at improving availability~\cite{menasche2013content}, does not work well: Bundling proactively increases availability by combining two or more items.
	%, . since all  items in the bundle are replicated on all users who request one of its constituent items. 
However, this creates larger downloads, increasing server traffic. For a majority of content item combinations, this increase is not offset by the decrease in server traffic resulting from additional availability of the content item. Even where it saves server traffic, the average delta gains are small, between 2\%--7\%.  
\section{Traffic Gain and Swarm Capacity: A Simple Analytical Model}
\label{sec:analytical_model}
In this section, we quantify the gains in terms of server traffic reduction from deploying hybrid peer-assisted content distribution for CDNs, and how it changes as the system scales. We first develop intuition for the savings in (edge-) server traffic in the context of a single content swarm by introducing a simple model which relates the gain to the number of users in the swarm, its \emph{capacity}. We then extend this to an expression for the gain across the entire corpus. Although we make simplifying assumptions for analytical tractability, \S\ref{sec:system} shows that the effects of various obstacle factors within our dataset agree well with the model, in the sense that given the decrease in peer upload capacity caused by a particular factor, the decrease in traffic gain as predicted by the model is observed in practice. \tref{tbl:params} lists the main parameters used in this and subsequent sections.
\begin{table}
\small{
\centering
\begin{tabular}{l|p{6cm}}
\toprule
Variable & Description\\
\midrule
$G$ & traffic gain from peer-assisted content delivery\\
$T_s$ & traffic between system's servers (or CDN edge servers) and clients' computers\\
$T_u$ & total amount of bytes watched in the system\\
$r_i$ & peer arrival rate of content $i$\\
$u_i$ & average session duration of content $i$\\
$c_i$ & capacity (i.e., average number of users) of content swarm $i$\\
$l_i$ & length of content $i$\\
$\beta_i$ & bitrate of content $i$\\
$E[B_i]$ & expected duration of availability period of content $i$\\
$P_i$ & unavailability probability of content $i$\\
$m$ & minimum number of online peers required to sustain a content swarm\\
\bottomrule
\end{tabular}
\caption{Parameters of the analytical model}\vspace{-4mm}
\label{tbl:params}}
\end{table}

%%%%%%%%%%%%%%%%%%%%%%%%%%%%%%%%%%%%%%%%%%%%%%%%%%%%%%%%%%%%%%%%%%%%%%%%%%%%%%%%%%%%%%%%%%
%%%%%%%%%%%%%%%%%%%%%%%%%%%%%%%%%%%%%%%%%%%%%%%%%%%%%%%%%%%%%%%%%%%%%%%%%%%%%%%%%%%%%%%%%%
\subsection{Traffic Gain}
We wish to understand the potential traffic savings from peer assistance, which we term as \emph{traffic gain}, or simply \emph{gain}. Formally, we denote with $T_s$ the total flow of client-server traffic in the system, i.e., the total amount of bytes transferred from system's servers (or CDN edge servers) to clients' computers, and with $T_u$ the total amount of content bytes watched by the users. In the case when a peer-assisted hybrid CDN strategy is deployed, the content can be delivered to a user either from a content delivery node (i.e., from a server) or from other users in the network (i.e., peers).  To measure the extent to which peers can offload traffic from the content provider or CDN's servers in a hybrid CDN setup, we define the \emph{traffic gain} metric: 
\begin{align}
G = 1 - \frac{T_s}{T_u}
\label{eq:cdn_gain}
\end{align}

Clearly the gain will be 0 in traditional content delivery when no peer assistance is exploited ($T_s = T_u$) and will be reaching values closer to $1$ for the ideal hybrid CDN where content access patterns are amenable to share content amongst peers. Note that $G$ can be negative in certain situations, for instance, if a server speculatively sends unrequested content items to peers in order to increase availability. This can occur in strategies which employ pre-fetching or push-based content delivery, or when content items are bundled together to increase availability.

%%%%%%%%%%%%%%%%%%%%%%%%%%%%%%%%%%%%%%%%%%%%%%%%%%%%%%%%%%%%%%%%%%%%%%%%%%%%%%%%%%%%%%%%%%
%%%%%%%%%%%%%%%%%%%%%%%%%%%%%%%%%%%%%%%%%%%%%%%%%%%%%%%%%%%%%%%%%%%%%%%%%%%%%%%%%%%%%%%%%%
\subsection{Swarm Capacity}
We wish to study how traffic gain evolves as the system scales. We use the average number of peers in the system to measure the scale of the system. We term this as the \emph{swarm capacity} or \emph{peer capacity}. With more users in the swarm, there are more peers to upload content to other peers, hence we also interchangeably use  the term \emph{peer upload capacity} or simply \emph{capacity}. 

Menasche \textsl{et al.}~\cite{menasche2013content} model  this self-scaling property of peer-to-peer swarms by treating each swarm as a queuing system with infinite servers: users who arrive at a swarm do not wait to be serviced, and can be served instantly by other members of the swarm. A user who arrives when the swarm is empty (or when there are too few peers to sustain swarming), departs immediately without being serviced by the swarm (In our case, this user is instead served by the edge servers of the CDN, and starts a new swarm). 

Consider a swarm $i$ for sharing a content item. Since there is no queuing time, the average time spent by users in the system is simply the average time spent watching the content, $u_i$. If users arrive at an average rate $r_i$, then according to Little's Law, the capacity can be written as 
\begin{align}
c_i = u_i r_i.
\label{eq:capacity}
\end{align}

%%%%%%%%%%%%%%%%%%%%%%%%%%%%%%%%%%%%%%%%%%%%%%%%%%%%%%%%%%%%%%%%%%%%%%%%%%%%%%%%%%%%%%%%%%
%%%%%%%%%%%%%%%%%%%%%%%%%%%%%%%%%%%%%%%%%%%%%%%%%%%%%%%%%%%%%%%%%%%%%%%%%%%%%%%%%%%%%%%%%%
\subsection{Relating Capacity to Traffic Gain}
For the simple case when content items are only downloaded from the server when they are unavailable among peers,  server traffic accounts for the portion of $T_u$ for which a sufficient number of peers were unavailable to sustain P2P delivery. Suppose $P_i$ is the probability that there are no users in the queuing system. If we assume that a minimum coverage of $m$ users is required to sustain a stable swarm (e.g., otherwise, a part of the file may become unavailable with high probability or the total upload bandwidth of seeding peers may not be sufficient to serve requesting peers), then define $P_i$ as the probability that the number of users drops below $m$.
Then we can write 
\begin{align}
T_s = P_i \times T_u,
\end{align} and the gain becomes:
\begin{align}
 G = 1 - P_i 
\label{eq:gain_main}
\end{align}  

%%%%%%%%%%%%%%%%%%%%%%%%%%%%%%%%%%%%%%%%%%%%%%%%%%%%%%%%%%%%%%%%%%%%%%%%%%%%%%%%%%%%%%%%%%
%%%%%%%%%%%%%%%%%%%%%%%%%%%%%%%%%%%%%%%%%%%%%%%%%%%%%%%%%%%%%%%%%%%%%%%%%%%%%%%%%%%%%%%%%%
%\subsection{Effect of different obstacle factors on gain}
  \begin{figure*}
    \centering
      \includegraphics[width=0.90\textwidth]{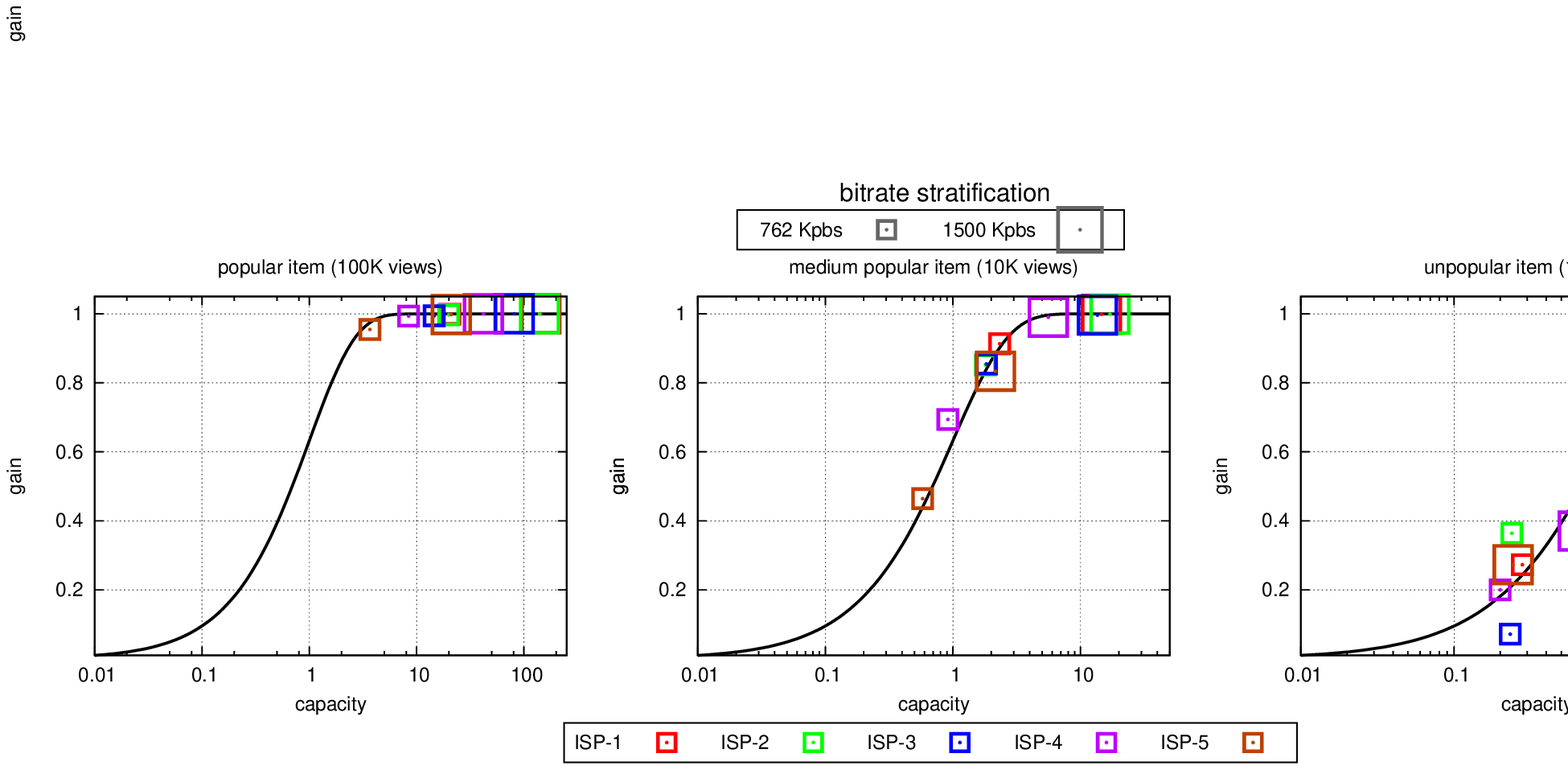}
        \vspace{-4mm}
      \includegraphics[width=0.90\textwidth]{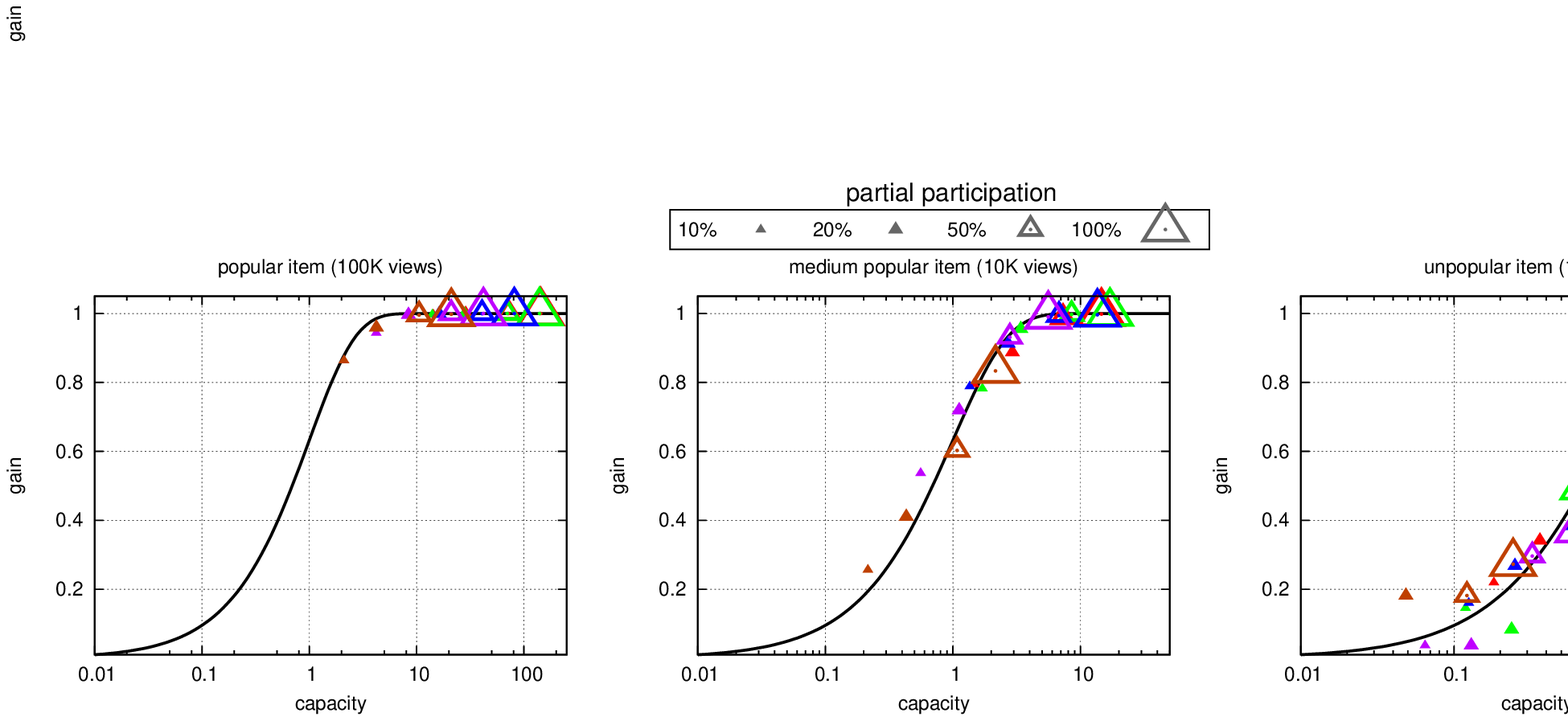}
        \vspace{-4mm}
      \includegraphics[width=0.30\linewidth]{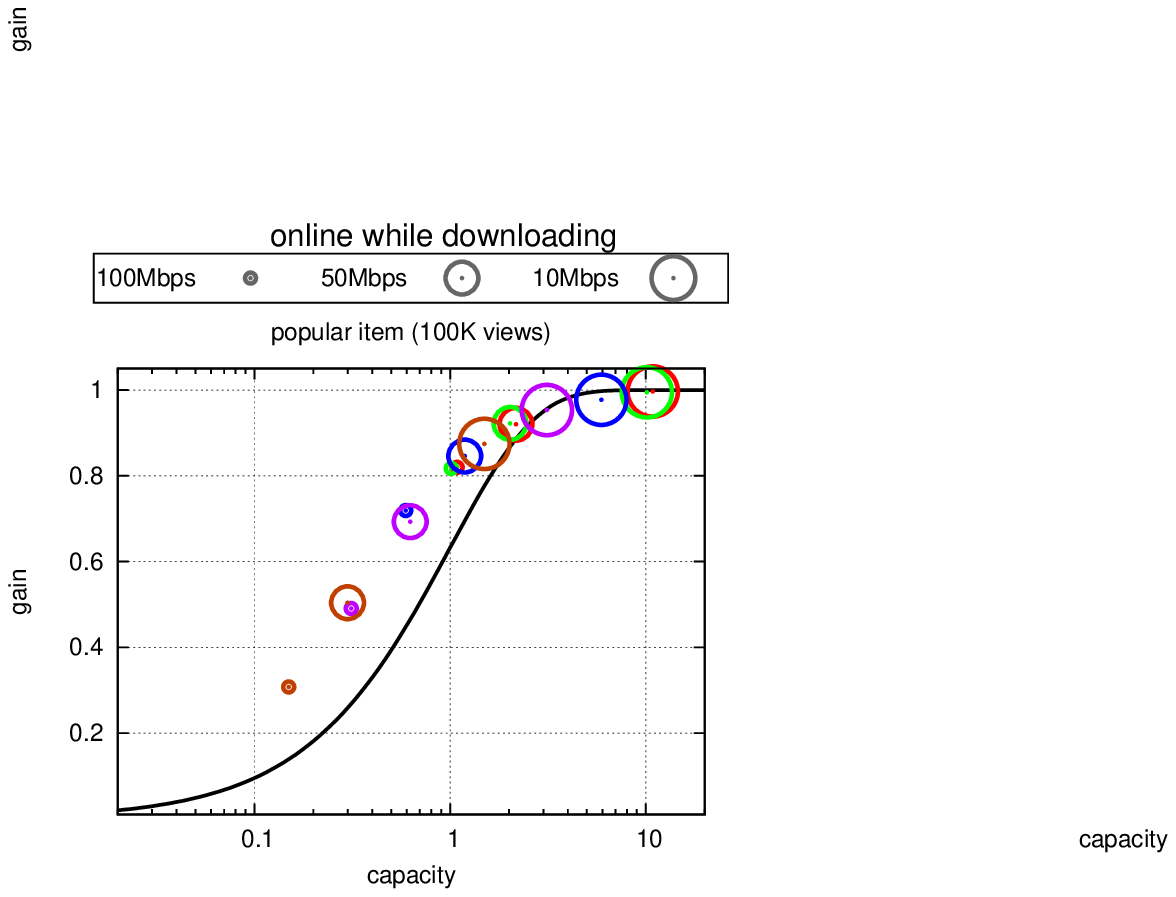}
      \includegraphics[width=0.30\linewidth]{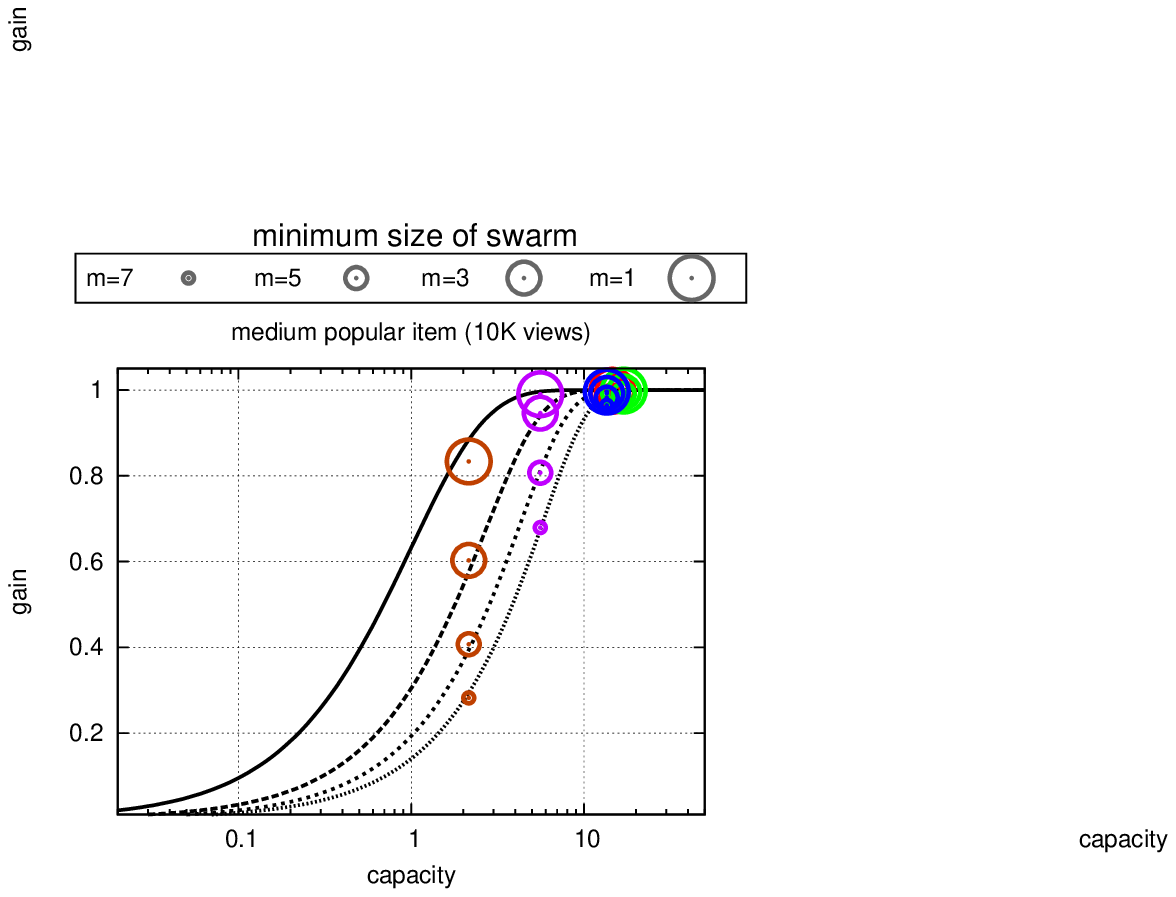}
      \includegraphics[width=0.30\linewidth]{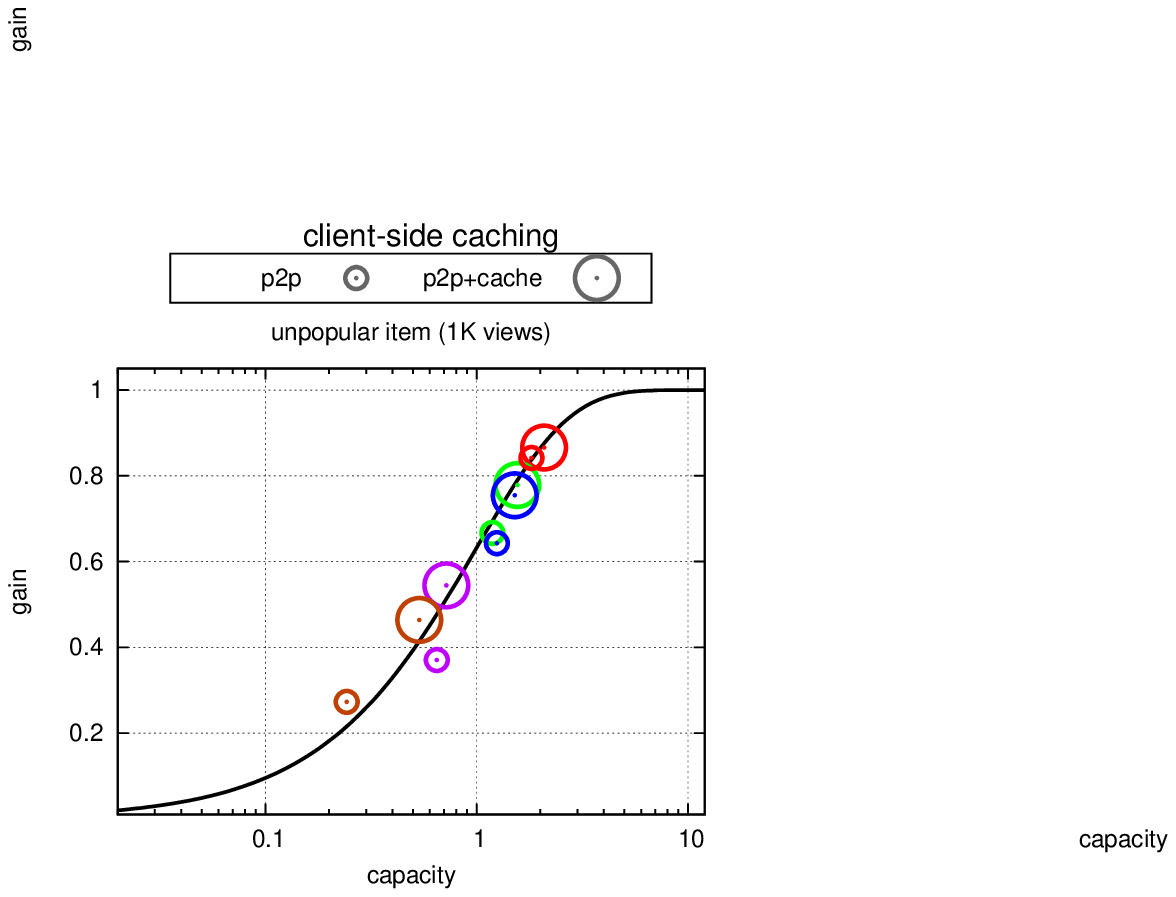}
      %\subfloat[popular item (100K views)\label{fig:popular}]
        %{\includegraphics[width=0.31\textwidth]{figures/single_item_partial_sharing_5.eps}}
      %\subfloat[medium popular item (10K views)\label{fig:medium}]
        %{\includegraphics[width=0.31\textwidth]{figures/single_item_partial_sharing_2.eps}}
      %\subfloat[unpopular item (1K views) \label{fig:unpopular}]
        %{\includegraphics[width=0.31\textwidth]{figures/single_item_partial_sharing_3.eps}}
        \vspace{-3mm}
    \caption{Traffic gains estimated theoretically (black curve) and via simulations (points), for exemplar highly popular (Left col.), medium popular (Centre col.) and unpopular (Right col.) content items, across top 5 ISPs (different colours). Top row shows effect of bitrate stratification; middle row, partial participation. Bottom row: Effect of---peers departing swarm after downloading at various bandwidths, popular item (Bottom row, Left); increasing minimum swarm size, medium popular item (Bottom row, centre); adding caching support to an unpopular item (Bottom row, right). Middle and bottom rows use swarms for 1500 Kbps rate. Bottom row assumes 100\% participation rate.}
        \vspace{-4mm}
  \label{fig:gain_individual_swarms}  
      \subfloat[Daily gains of the entire content corpus \label{fig:gain_temporal}]
        {\includegraphics[width=0.29\textwidth]{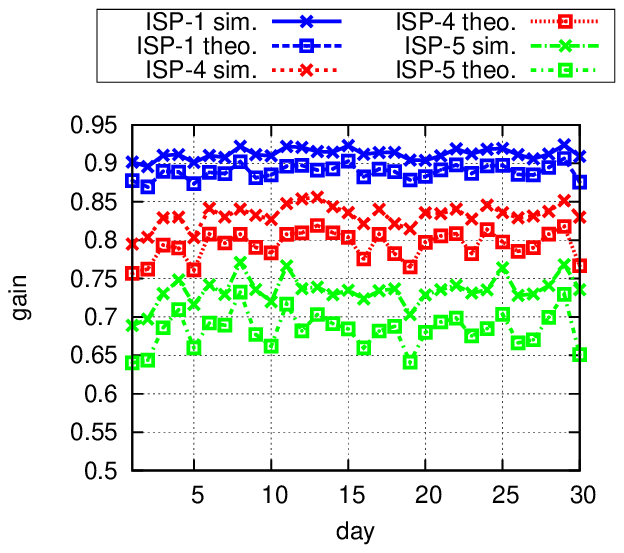}}\hspace{1em}
      \subfloat[Average gains with partial participation \label{fig:gain_pp}]
        {\includegraphics[width=0.29\textwidth]{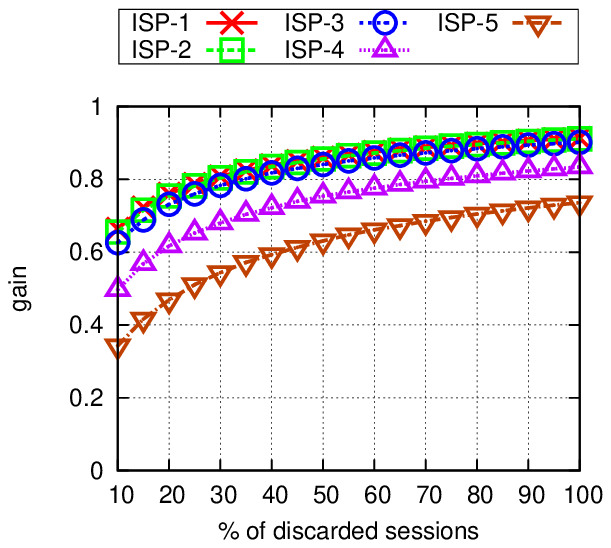}}\hspace{1em}
      \subfloat[Average gains with constraints on the minimum number of peers in swarms \label{fig:gain_limited}]
        {\includegraphics[width=0.29\textwidth]{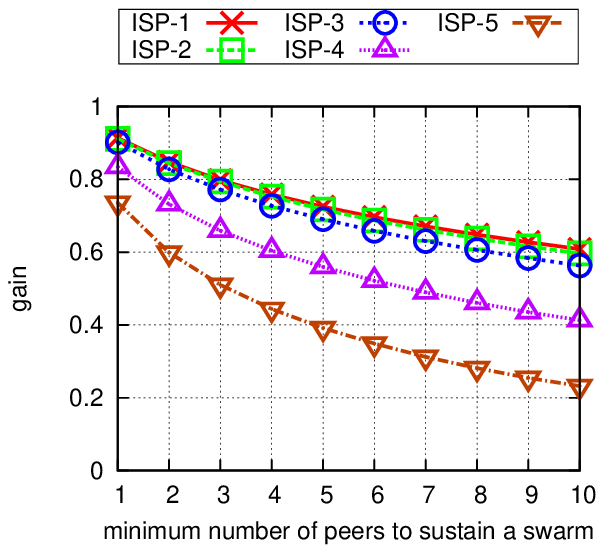}}
        \vspace{-2mm}
    \caption{The aggregate traffic gain for the entire corpus, across various ISPs throughout the month of Sep 2013.}
    \label{fig:gain-aggregate}
  \end{figure*}

For analytical tractability, we  follow Menasche \textsl{et al.} and consider the infinite server queuing system as an $M/G/\infty$ queue. 
%i.e., clients' requests for a content item $i$ are arriving according to a Poisson process with rate $r_i$\footnote{\nn{Although we observe diurnal patterns in our data, arrivals over small periods of time (e.g., 3--4 hour windows) appear to fit the Poisson assumption (according to a Chi-Square test with T=5\% significance) for about 40\% of content items. Among the items which do not fit the assumption, about30\%?? do not receive enough arrivals for a statistically significant test; others appear to have a different arrival pattern.}}. 
It then follows from standard $M/G/\infty$ results that:
\begin{align}
 P_i = \frac{1}{E[B_i]r_i + 1}
\label{eq:probability_of_unavailability}
\end{align}
where $E[B_i]$ is the expected duration of time periods during which the content is available among the peers. 
Clearly, $E[B_i]$ depends on the arrival rate $r_i$ and the length of time intervals $u^j_i$ during which users stay online. This relation, although not trivial for the general case, can be expressed with a closed-form solution for some specific distributions of $u^j_i$. In this current work, we assume that the time intervals during which users stay online while watching content $i$, are exponentially distributed with expectation $u_i$. 
%We validate this assumption via realistic traces in the next session. 
We can then employ ~\cite[Lemma 3.3]{menasche2013content} to derive an expression for $E[B_i]$:
\begin{align}
 E[B_i] = \frac{u_i (1 + e^{u_i r_i}(u_i r_i)^{-m} (m \Gamma(m) - \Gamma(1+m,u_i r_i)))}{m}
% E[B_i] = \frac{u_i (1 + \frac{e^{u_i r_i}}{(u_i r_i)^{-m}} (m \Gamma(m) - \Gamma(1+m,u_i r_i)))}{m}
% E[B_i]r_i + 1= \frac{u_i}{m} + u_i \sum_{j=1}^{\infty}{(u_i r_i)^j \frac{(m-1)!}{(m+j)!}}
\label{eq:expected_busy_period}
\end{align}
where $m$ is the minimum number of peers needed to sustain a swarm, and $\Gamma(x)$ and $\Gamma(x,y)$ are Gamma and incomplete Gamma function, correspondingly. 
Substituting Eq.~\eqref{eq:expected_busy_period} into Eq.~\eqref{eq:probability_of_unavailability}, and using Eq.~\ref{eq:capacity}, we can write $G$ as a function of $c_i$ and $m$:
\begin{align}
 G = 1 - (\frac{c_i (1 + e^{c_i}c_i^{-m} (m \Gamma(m) - \Gamma(1+m,c_i)))}{m} + 1)^{-1}
% G = \frac{u_i (1 + \frac{e^{u_i r_i}}{(u_i r_i)^{-m}} (m \Gamma(m) - \Gamma(1+m,u_i r_i)))}{m}
\label{eq:g_as_func_m_c}
\end{align}

%\subsection{Relating Capacity to Server Traffic}
Next, users interested in downloading content item $i$ at bitrate $\beta_i$ generate a traffic of $T_u = \beta_i l_i r_i$, where $l_i$ is the length of the item $i$ and $r_i$ is the peer arrival rate as above. Thus, when peer assistance is employed, the server traffic generated is
\begin{align}
 T_s &= \beta_i l_i r_i P_i \notag\\
     &= \beta_i l_i (\frac{c_i (1 + e^{c_i}c_i^{-m} (m \Gamma(m) - \Gamma(1+m,c_i)))}{m} + 1)^{-1}
\label{eq:server_traffic}
\end{align}  

%%%%%%%%%%%%%%%%%%%%%%%%%%%%%%%%%%%%%%%%%%%%%%%%%%%%%%%%%%%%%%%%%%%%%%%%%%%%%%%%%%%%%%%%%%
%%%%%%%%%%%%%%%%%%%%%%%%%%%%%%%%%%%%%%%%%%%%%%%%%%%%%%%%%%%%%%%%%%%%%%%%%%%%%%%%%%%%%%%%%%
\subsection{Intuition for Single Swarm Performance}
To drive intuition, we consider the simple case when only one other peer is required to sustain the swarm (i.e., m = 1). In this case, we obtain the simple relation $G=1-e^{-c_i}$. Thus gain improves exponentially with capacity.

Similarly, when $m=1$, we get $T_s = \beta_i r_i l_i e^{-c_i}$ which has a maximum at $c_i = 1$. Thus, server traffic initially increases as offered load, in terms of number of peers requesting the item, increases. It reaches a maximum around when the swarm becomes self-sustaining, with $c_i=1$ user on average within the system. Any subsequent increase in peer numbers decreases server traffic as the swarm takes over the load.

%%%%%%%%%%%%%%%%%%%%%%%%%%%%%%%%%%%%%%%%%%%%%%%%%%%%%%%%%%%%%%%%%%%%%%%%%%%%%%%%%%%%%%%%%%
%%%%%%%%%%%%%%%%%%%%%%%%%%%%%%%%%%%%%%%%%%%%%%%%%%%%%%%%%%%%%%%%%%%%%%%%%%%%%%%%%%%%%%%%%%
\subsection{Traffic Gain Across Multiple Swarms}
It is straightforward to measure gain across multiple swarms analytically by taking the sum of server traffic $T^i_s$ and total traffic $T^i_u$ across all swarms $i$ being considered: 
\begin{align}
 G = 1 - \frac{\sum_{i}[\frac{\beta_i l_i r_i}{E[B_i]r_i + 1}]}{\sum_i[\beta_i l_i r_i]}
\label{eq:gain_multi_swarm}
\end{align}   

This can not only be used to get the gain across the entire content corpus, but also to measure the effect of different obstacle factors: When there are no obstacle factors considered, there is one swarm per content item.  ISP-friendliness and bitrate stratification create smaller swarms, one for each bitrate, and one for each ISP. The smaller swarms have lower rates of arrivals $r_i^1, r_i^2,\ldots$, such that $r_i = r_i^1 + r_i^2+\ldots$. 

The case of partial participation is slightly harder: only a fraction $\alpha$ of  peers are willing to upload, or equivalently,  the ratio of upload to download bandwidths of peers is $\alpha$. This decreases availability (or increases unavailability probability to a value $P_i'$) equivalent to decreasing the rate of arrivals \emph{for peer uploads} to $r'_i = \alpha \times r_i$. However, the actual rate of arrivals still remains $r_i$. Thus, we have  a server traffic of 
\begin{align}
T'_s = \sum_{i}{[\beta_i l_i r_i P'_i]} = \sum_{i}{[\frac{\beta_i l_i r_i}{E[B'_i] r_i \alpha + 1}]}
\label{eq:server_traffic_partial_participation}
\end{align}   
where $B'$ is the expected duration of availability periods for the swarm $i$ with limited participation, and can be calculated by substituting arrival rate with $r_i'$ in Equation~\ref{eq:expected_busy_period}. Hence, the gain for the partial participation case takes the form
\begin{align}
G = 1 - \frac{T'_s}{T_u} = 1 - \frac{\sum_{i}{[\frac{\beta_i l_i r_i}{E[B'_i] r_i \alpha + 1}]}}{\sum_{i}{[\beta_i l_i r_i]}}
\label{eq:gain_main_partial_participation}
\end{align}   
We use Eq.~\ref{eq:gain_main_partial_participation} in the following section to assess the aggregated traffic gain ($G_{theo}$) in multi-swarm systems and validate by comparing the corresponding results from simulations ($G_{sim}$).

%%%%%%%%%%%%%%%%%%%%%%%%%%%%%%%%%%%%%%%%%%%%%%%%%%%%%%%%%%%%%%%%%%%%%%%%%%%%%%%%%%%%%%%%%%
%%%%%%%%%%%%%%%%%%%%%%%%%%%%%%%%%%%%%%%%%%%%%%%%%%%%%%%%%%%%%%%%%%%%%%%%%%%%%%%%%%%%%%%%%%
\section{Empirical Analysis}
\label{sec:system}
%\section{Multi-Swarm Systems} 

%Here we consider the aggregated traffic gain in a multi-swarm system. Our goal is to assess the contributions of individual content swarms on the system in whole.  

In this section, we empirically analyse  traffic gains in our large trace, to complement the intuition derived from the model of the previous section. We  study the effect of various obstacle factors under workload paramaters derived from the trace. 

\begin{figure}
\subfloat[Share of traffic per content item\label{fig:contentpop}]
			{\includegraphics[width=0.48\columnwidth]{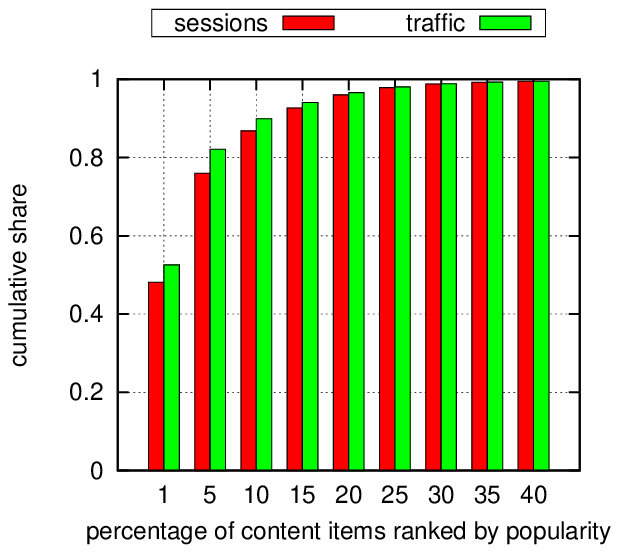}}\hspace{1em}
 \subfloat[Temporal decay of number of views per day across content items\label{fig:freshness}]
			{\includegraphics[width=0.48\columnwidth]{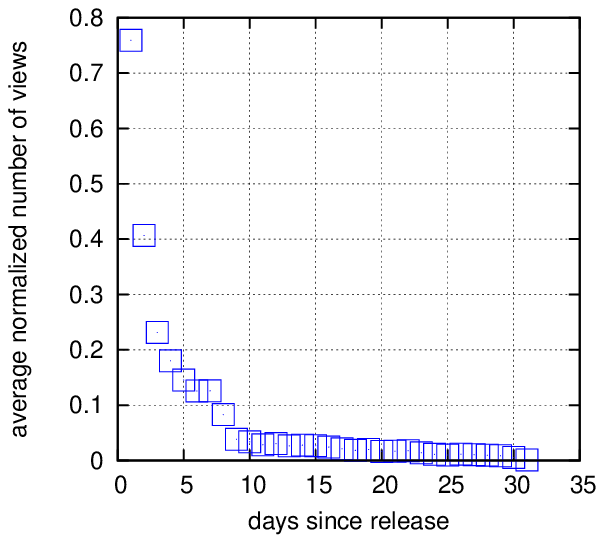}}
\caption{Popularity concentration: Many users are interested in the same items at the same time.}\vspace{-4mm}
\end{figure}
%%%%%%%%%%%%%%%%%%%%%%%%%%%%%%%%%%%%%%%%%%%%%%%%%%%%%%%%%%%%%%%%%%%%%%%%%%%%%%%%%%%%%%%%%%
%%%%%%%%%%%%%%%%%%%%%%%%%%%%%%%%%%%%%%%%%%%%%%%%%%%%%%%%%%%%%%%%%%%%%%%%%%%%%%%%%%%%%%%%%%
\subsection{Simulation details}
We implemented an event-driven simulator where timestamps of events, i.e., start times and durations of user sessions, are taken from the BBC iPlayer trace. Our trace also provides the bitrates for each session. Peers are assigned to swarms based on their ISP and bitrate\footnote{Later in this section we show that bitrates are likely to remain stable during individual sessions and typically feature values very close to the corresponding maximum bitrate values. Therefore, in order to match bitrate requirements of individual peers, we map average per-session bitrates to the closest out of nine different maximum bitrates available in iPlayer and split content swarms accordingly.}. On each simulation step, we analyze the number of available peers in the network for the content item being requested and make a decision on whether the user session being processed can be served by other peers or from the server. We conservatively estimate that a peer can be served by the swarm of other peers accessing the same content item at the same bitrate from the same ISP, if we find another concurrent user session in the swarm who has been streaming for at least 10\% of duration of the content item. This threshold ensures that the serving peer can buffer sufficient amount of content to satisfy the immediate streaming requirements of the receiving peer (with an average download bandwidth of $b = 18.7$ Mpbs~\cite{ofcom2014speed} the full length of a content item with bitrate $\beta = 762$ Kbps or $1500$ Kpbs can be buffered by users in the time required for watching the first $\beta / b \approx 4$\% or $8$\% of a content item). 
Note that we assume peer assignments can be managed centrally, similar to NetSession~\cite{zhao2013peer} and other managed swarming techniques~\cite{peterson2009antfarm}. %Thus, in contrast with BitTorrent~\cite{cohen2008bittorrent}, users do not have to find peers themselves. 
For calculating the availability of peers, we also assume an ``online while you watch model'', i.e., that the content is available for upload from a peer if that peer is currently watching the content\footnote{Unlike ``selfish'' file download techniques like BitTorrent~\cite{cohen2008bittorrent}, ensuring peer availability whilst a video is being watched is straightforward in a streaming  model. 
%Peers can buffer any amount of content ahead of playback, to make use of instantaneous bandwidth availability. However, 
Content buffered on the client can be protected by DRM, and securely decoded by the video player at playback time only upon receiving permission from the server. A simple mechanism, such as requiring regular heartbeats back to the server in order for the player to show the video, can be used to ensure that the user stays online and shares content while she watches the video. Peers receiving content from this user can also independently verify to the server that content is being shared, in order for the player to obtain permission from the server to unlock the content.}. %Again, this is in direct contrast to ``selfish'' protocols like BitTorrent~\cite{cohen2008bittorrent}, where peers can disappear immediately after downloading. %In principle, we can allow peers to buffer bytes ahead of playback time and fetch content as fast as they can, to make use of instantaneous bandwidth which may be available. However, we require them to stay online\footnote{We assume that content fetched ahead of playback can be protected using Digital Rights Management by iPlayer. The current BBC iPlayer app allows users the option to download entire programmes, and securely play it back offline, using similar mechanisms (these are not included in our traces). A simple mechanism such as requiring regular heartbeats back to the server in order for the player to show the video, can be used to ensure that the user stays online and shares content.}. 
% One simplification we make is that users who have watched 10\% of a content item have the full length $l_i$ downloaded. We also remove all sessions where a user has watched less than $10$\% of the content from the analysis. 
%Note that we do not take BitTorrent-like chunking  and incentives~\cite{cohen2008bittorrent} into consideration. This is not necessary because, in peer-assisted swarming similar to , can be used to directly match peers and allocate their bandwidths, removing the need for explicit incentives and co-operation.

\begin{figure}
%\subfloat[share of traffic per ISP \label{fig:ispsplit}]
%      {\includegraphics[width=0.40\textwidth]{figures/isp_barcharts.eps}} 
\centering
      \includegraphics[width=0.65\columnwidth]{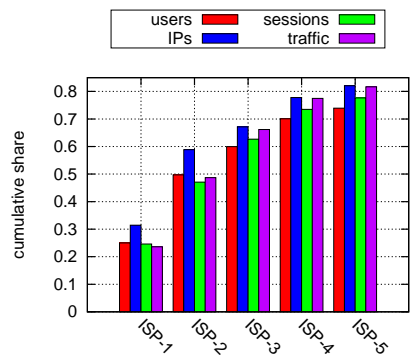}
      \caption{Share of traffic per ISP}\vspace{-4mm}
      \label{fig:ispsplit}
\end{figure}

In this simulation framework, we can calculate useful traffic $T_u$ and server $T_s$ traffic generated in the network in our simulation, and apply Equation~\ref{eq:cdn_gain} to calculate the gain according to simulation, $G_{sim}$. To account for daily patterns in users' activity we ran simulations for individual days
%, assuming that a day starts at 3AM when the user activity is minimal, 
and compare the results with the theoretical estimations calculated from Equation~\ref{eq:g_as_func_m_c} (single swarm) and Equation~\ref{eq:gain_multi_swarm} (multiple swarms), i.e., $G_{theo}$. Finally, we use Equation~\ref{eq:capacity} to calculate capacity $c_i$ of individual content swarms. 
%The goal of this analysis is to assess the extent to which the proposed analytical model can capture temporal patterns of user activity emerging from our workload.

%%%%%%%%%%%%%%%%%%%%%%%%%%%%%%%%%%%%%%%%%%%%%%%%%%%%%%%%%%%%%%%%%%%%%%%%%%%%%%%%%%%%%%%%%%
%%%%%%%%%%%%%%%%%%%%%%%%%%%%%%%%%%%%%%%%%%%%%%%%%%%%%%%%%%%%%%%%%%%%%%%%%%%%%%%%%%%%%%%%%%
\subsection{Effect of different obstacle factors on individual swarms}

Figure~\ref{fig:gain_individual_swarms} shows the effect of different obstacle factors on swarms of different sizes. Specifically, we consider three different content items with various levels of popularity and hence swarm sizes: an episode of the highly popular ``Bad Education'' series which accounts for over $100$K views in September 2013 (Left column) and episodes of a medium popular item, "Question Time" (Centre column), and an unpopular item, "What's to Eat"  (right column) with around $10$k and $1$k views, respectively.  We analyze the gain of various content swarms as a function of their capacity and measure the traffic gains in ISP-friendly swarms, when peer-to-peer traffic is localized inside ISPs. The top-5 ISPs by number of sessions  are considered. Within each  ISP, peers  watching at different bitrates are separated from each other (Top row). We also consider various levels of peers' participation, when only a portion of peers participate in peer-to-peer content distribution (middle row).

Focusing first on the \textbf{top row}: The highly popular item is hardly affected by the obstacle factors, and gains remain consistently high across all ISPs and bitrates--well over 90\% of traffic is saved as the capacity of the swarm remains high even after taking the obstacle factors into account. However, less popular items are affected to a greater extent. As we will see later, 1500Kbps is the more popular than 762 Kbps; thus gains are smaller for the smaller sized-swarms of 762 Kbps, dropping down to less than 20\% gain for the unpopular item. 

Next the \textbf{middle row}: Observe in general that the gains are lesser in the middle row than the top row, as participation levels of less than 100\% are considered. Also, observe that again, the unpopular and medium popular items are affected more than the popular item, where gains remain above 80\% even with only 10\% of peers participating.

The \textbf{bottom left} panel of Figure~\ref{fig:gain_individual_swarms} shows that the ``online while you watch'' assumption is critical to the high gains observed. Recall that we assume that users remain available to upload content as long as they are watching content. Given the relatively long duration of TV shows, we expect this would greatly increase content availability and hence swarming capacity. To test this assumption, we consider the effect of an ``online while downloading'' scenario, of peers being able to download at a bandwidth $b_i$ which is higher than the bitrate $\beta_i$, and then departing the swarm immediately after download. In other words, we consider the effect of peers uploading content only for the duration $\min(u_i,\beta_i * l_i / b_i)$ rather than for their entire period of watching the show, $u_i$. The bandwidths $b_i$ we consider (10Mbps, 50Mbps and 100Mbps) are all realistic given current broadband rates in the UK~\cite{ofcom2014speed}. The importance of the ``online while you watch'' assumption can be seen from the fact that even for the extremely popular item considered in the bottom left panel, gains from peer assistance can more than halve, and drop to nearly 30\% for a top-5 ISP.

\begin{table}
\small{
 \begin{center}
		\begin{tabular}{|l|l|l||l|l|}
			\cline{2-5} 
			\multicolumn{1}{l|}{} & \multicolumn{2}{c||}{No stratification} & \multicolumn{2}{c|}{Stratification}\\ \hline
			\textbf{ISP} & \textbf{$G_{sim}$}  
			& \textbf{$G_{theo}$} & \textbf{$G_{sim}$} & \textbf{$G_{theo}$} \\ \hline
			ISP-1 & 0.92 & 0.90 & 0.88 & 0.84 \\ \hline
			ISP-2 & 0.92 & 0.89 & 0.87 & 0.84 \\ \hline
			ISP-3 & 0.91 & 0.87 & 0.86 & 0.81 \\ \hline
			ISP-4 & 0.86 & 0.81 & 0.78 & 0.72 \\ \hline
			ISP-5 & 0.77 & 0.71 & 0.67 & 0.60 \\ \hline   

		\end{tabular} 
	\end{center}
	\caption{Traffic gain across various ISPs with and without bitrate stratification}
	\label{tab:cd_gain_multi_swarm}}
\end{table}
	
The \textbf{middle panel of the bottom row} in Figure~\ref{fig:gain_individual_swarms} summarises the effect of requiring a minimum swarm size $m$ for the medium popular item (similar effects are seen for other items as well). This is a simplified test to account for the known fact that small swarms are less stable~\cite{liu2009peers}: In small swarms, peer departure may permanently remove some parts of the content item, requiring server assistance to complete downloads. When we impose constraints on the minimum number of peers in the system, gain quickly drops considerably. %This occurs because average swarm sizes during night and early morning hours can be quite small, 

Finally, we note that simulation results across all experiments are in good agreement with theoretical estimations computed with Eq.~\ref{eq:g_as_func_m_c} (i.e., black curve in plots) in all cases. 

%Clearly, constraints imposed by either traffic localization or partial participation of peers are leading to a reduction of a swarm's capacity, where the capacity of $u \times r = 1$ (i.e., one peer arriving during an average user's session) is enough to save up to $65$\% of traffic. Particularly, for a swarm of a very popular content item with the request rate $89$ times higher than the average session duration (i.e., $u \times r = 89$), as much as $10$\% are required to sustain a remarkably high traffic gain of $86$\%. The gain is less considerable for an unpopular item with the capacity $u \times r < 2$, when only $35-75$\% (\dk{check these numbers as the plot has been updated}) can be saved even if all $100$\% of peers are participating in the peer-assistance. Note, however, that the traffic $T_u$ generated by a content swarm is growing proportional to its capacity, therefore, high savings from popular content swarms may potentially compensate low gains from unpopular ones. We will analyze this tradeoff between the size of swarms and their contribution in the aggregated traffic gain in the following section. 
 
%%%%%%%%%%%%%%%%%%%%%%%%%%%%%%%%%%%%%%%%%%%%%%%%%%%%%%%%%%%%%%%%%%%%%%%%%%%%%%%%%%%%%%%%%%
%%%%%%%%%%%%%%%%%%%%%%%%%%%%%%%%%%%%%%%%%%%%%%%%%%%%%%%%%%%%%%%%%%%%%%%%%%%%%%%%%%%%%%%%%%
\subsection{Aggregate traffic gain across  corpus}
\label{sec:aggregate_results}
Having explored the space of obtainable gains with  three exemplar content items, we next present the aggregate gain for all items in the content corpus in Figure~\ref{fig:gain-aggregate}. As we examine each obstacle factor, we also attempt to explain the  high gains we see, in terms of the characteristics of the content corpus. 

To start off, we observe that, as common in most content corpora, there is a huge popularity skew, with the top 10\% of items accounting for over 80\% of traffic and sessions (\fref{fig:contentpop}). Additionally, a majority of accesses happen just after a content item has been released (\fref{fig:freshness}). Thus, although it is an on-demand workload, we expect  users to be interested in the same popular items around the same time. These conditions are conducive for large swarming capacity and high gains.

\fref{fig:gain_temporal} shows the average daily gains across all items in the content corpus. Gains remain high ($>50\%$) for each of the top-5 ISPs we consider, even after splitting by ISP\footnote{Experiments in \fref{fig:gain-aggregate} are for sessions in the most popular bitrate, i.e., 1500 Kpbs.}. As shown in \fref{fig:ispsplit}, these ISPs together account for over 70\% share by any measure (users, IPs, sessions, traffic), thus the majority of system-wide gains are captured here. Also, observe that the top two ISPs have a $\approx$24\% traffic share each, and the next two 17\% and 11\% traffic shares, correspondingly. This split again creates conditions for relatively large swarms, helping keep gains high. 

%We observe in passing that gain improves on Wednesday and drops on Thursdays and Mondays. This pattern reflects a common trend in users' activity when thousands of users catch-up with a new episode of the popular TV show "Bad Education", usually released on Wednesdays.

\begin{figure}
\vspace{-8mm}
\subfloat[share of traffic per bitrate\label{fig:bitrates}]{\includegraphics[width=0.48\columnwidth]{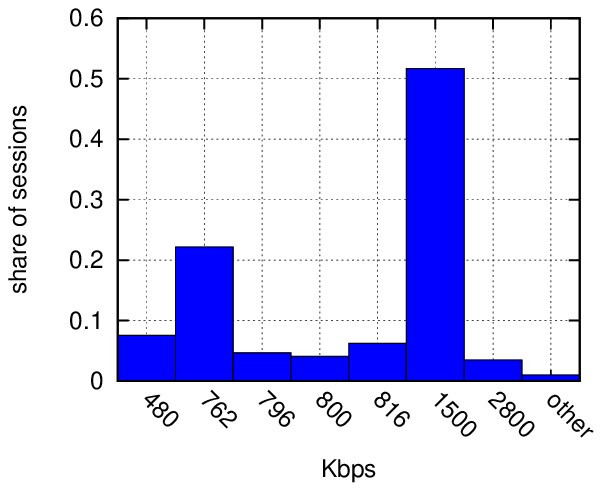}}\hspace{1em}
\subfloat[deviations between max, min and average bitrates in users' sessions \label{fig:bitrates_deviation}]{\includegraphics[width=0.48\columnwidth]{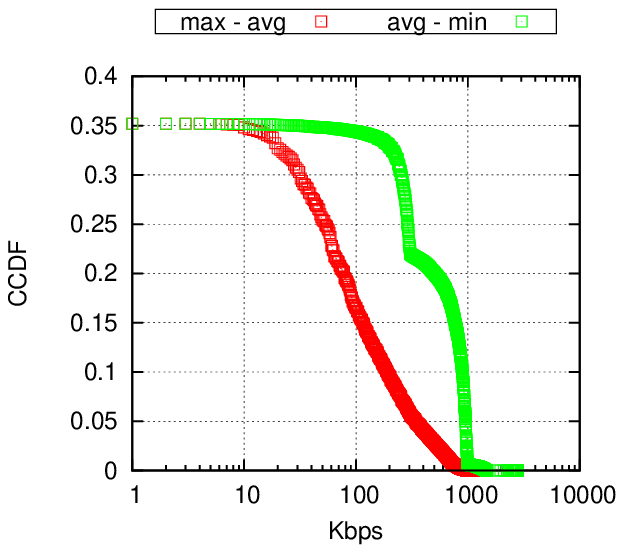}}
\caption{Bitrate characteristics across user sessions}\vspace{-4mm}
\label{fig:bitrate_char}
\end{figure}

Next, in Table~\ref{tab:cd_gain_multi_swarm} we present the results of bitrate stratification across various IPSs. As with ISP friendliness, we note a remarkably high gain for the aggregated traffic across all considered ISPs. To explain this, we turn to \fref{fig:bitrate_char}. As shown in \fref{fig:bitrates}, two bitrates dominate, collectively accounting for over 70\% of sessions, and suggesting that rendering bitrate stratification is ineffective in decreasing swarm capacity: most of the user sessions belong to the top 2 bitrates, keeping swarm sizes high. Remarkably, we also observe that bitrates do not change often during a session. \fref{fig:bitrates_deviation} shows the Complementary Cumulative Distribution (CCDF) of the difference between maximum and average bitrates, and average and minimum bitrates, showing that only 35\% have a non-zero difference; thus for 65\% of sessions, average bitrate = min bitrate = max bitrate. Even when there is a difference, the average bitrate is closer to the maximum compared to the minimum. We conjecture that this is a result of the relatively high download rates possible in today's broadband networks, in comparison to the bitrates required for streaming (2800 Kbps is the maximum bitrate encoding used in iPlayer, compared to average residential speed of 18.7 Mbps~\cite{ofcom2014speed}). Thus, swarms split by bitrate have a high likelihood of remaining stable, which can also explain the ineffectiveness of bitrate stratification.

Finally, as already shown for individual content items, partial participation (\fref{fig:gain_pp}) and minimum swarm size requirements (\fref{fig:gain_limited}) have a large impact on corpus-wide gain in comparison with ISP-friendliness and bitrate stratification. 

%%%%%%%%%%%%%%%%%%%%%%%%%%%%%%%%%%%%%%%%%%%%%%%%%%%%%%%%%%%%%%%%%%%%%%%%%%%%%%%%%%%%%%%%%%
%%%%%%%%%%%%%%%%%%%%%%%%%%%%%%%%%%%%%%%%%%%%%%%%%%%%%%%%%%%%%%%%%%%%%%%%%%%%%%%%%%%%%%%%%%
\subsection{Implications and Notes about Generality}
We observed, both at an individual content item level, and at a system level, that relatively high traffic gains are possible in BBC iPlayer even if peer assisted content delivery is required to be ISP-friendly. We believe this result could extend to other countries, where similar dominance of a handful of ISPs has been observed (e.g., Verizon and Comcast in the USA). Similarly, the ineffectiveness of bitrate stratification and the relatively infrequent changes within a session, although initially surprising to us, may also extend to other settings where residential ISPs are relatively well provisioned in comparison to the needs of streaming websites. This paints an encouraging picture for the potential of peer-assisted streaming of on-demand, long-duration content. At the same time, the difficulties with partial participation and minimum swarm size requirements indicate the need for further improving content availability, which we examine next. 

\section{Content Bundling}

% 	\begin{figure*}
%     \centering
% 			\subfloat[content parameters in a bundle of two content items with positive traffic savings\label{fig:bundling_schema}]
%         {\includegraphics[width=0.23\textwidth]{figures/bundle_function.eps}}\hspace{2em}
% 			\subfloat[share of bundles with positive traffic savings \label{fig:bundling_share}]
%         {\includegraphics[width=0.25\textwidth]{figures/isps_bundles_updated.eps}}\hspace{2em}
% 			\subfloat[average traffic gains across content bundles\label{fig:bundling_gain}]
%         {\includegraphics[width=0.25\textwidth]{figures/isps_bundles_gain_updated.eps}}
%     \caption{Impact of content bundling on the traffic gains for a single bundle and for the system in whole}\vspace{-4mm}
%     \label{gain_two_content_bundle}
%   \end{figure*}

 	\begin{figure}
     \centering
 			\subfloat[share of bundles with positive traffic savings \label{fig:bundling_share}]
         {\includegraphics[width=0.47\columnwidth]{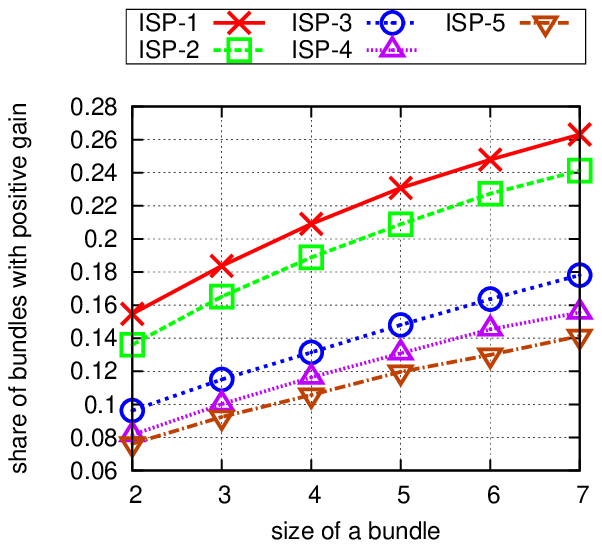}}\hspace{1em}
 			\subfloat[average traffic gains across content bundles\label{fig:bundling_gain}]
         {\includegraphics[width=0.47\columnwidth]{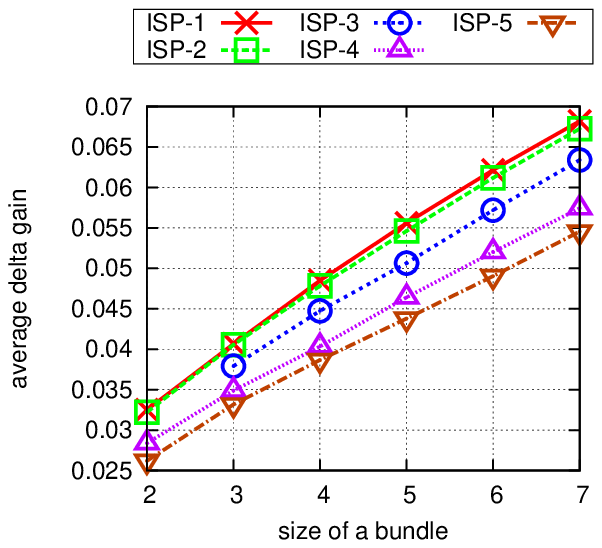}}
     \caption{Impact of content bundling on the traffic gains for a single bundle and for the system in whole}\vspace{-4mm}
     \label{gain_two_content_bundle}
   \end{figure}

Bundling can increase content availability. This idea has been actively discussed in the recent peer-to-peer literature~\cite{menasche2013content, lev2010dynamic, zhang2012dynamic, carlsson2010using, carlsson2012tradeoffs}. With bundling, individual content swarms are combined in larger bundle swarms, therefore, increasing the chance that bundled content is available among peers throughout the day. The impact of content bundling on traffic gains in peer-assisted hybrid CDNs is, though, unclear. Intuitively, the size of a bundle grows as more content items are added to it, therefore, inducing a \emph{traffic overhead} for each download from the server (i.e., when content bundle is not available among peers). In the extreme, a large bundle of unpopular content items is never available among peers yielding a traffic overhead proportional to the total weight of bundled items for each server request. In this section we study the tradeoff between these two factors (i.e., the increased availability and the server traffic overhead induced by content bundling) and assess their impact on the overall traffic savings. 

\subsection{Analytical model for bundles}

Formally, the aggregate arrival rate $R^b$ and the \emph{weight} (the total number of bytes) $\Omega^b$ of a content bundle grow as the sum of  the arrival rates and weights  of individual content items, i.e., $\Omega^b = \sum_{i = 1}^k{\beta_i l_i}$ and $R^b = \sum_{i = 1}^k{r_i}$, where $k$ is the \emph{size} of a bundle, i.e., the number of items in it. In contrast, the probability that a bundle is not available among peers (i.e., unavailability probability), decreases as the product of the corresponding probabilities $P_i$ for individual items of which it consists\footnote{we assume that availability of various content items at a given time are pairwise independent events and that each item belongs to only one bundle.}, i.e., $P^b = \Pi_{i = 1}^k{P_i}$.  %\frac{1}{e^{\sum_u{r_i u_i}}}$.
%~\footnote{We consider the case when no constraints on the upload bandwidth are imposed since our aim is to define the upper bound for the gain in the bundling scenario. It is worth nothing that by adding constraints the performance of bundling will drop as the content becomes heavier and more available content peers will be required to support peer-to-peer streaming.}. 
Then, server traffic $T^b_s$ generated by a bundled swarm $b$ can be calculated as a product of these three components, i.e.:
 
\begin{align}
 T^b_s = \Omega^b R^b P^b = \sum_{i = 1}^k{[\beta_i l_i]} \sum_{i = 1}^k{[r_i]} \Pi_{i = 1}^k{[P_i]}
\label{eq:server_traffic_bundle}
\end{align}   

To assess traffic savings from a single bundle we compare the server traffic $T^b_s$ with the total server traffic generated by individual content swarms without bundling, i.e.:  

\begin{align}
\Delta T_s &= \sum_{i = 1}^k{T^i_s} - T^b_s \notag\\ 
					 &= \sum_{i = 1}^k{[\beta_i l_i r_i P_i]} - \sum_{i = 1}^k{[\beta_i l_i]} \sum_{i = 1}^k{[r_i]} \Pi_{i = 1}^k{[P_i]}
\label{eq:delta_gain_single_bundle}
\end{align}   

Finally, we measure the gain in the system when bundling is enabled ($G^b$) and compare it with the benchmark result when no bundling is used ($G$). Formally we get:

\begin{align}
\Delta G = G - G^b = \frac{\Delta T_s}{T_u} 
\label{eq:delta_gain_bundling}
\end{align}   

where $T_u = \sum_{i = 1}^k{\beta_i l_i r_i}$ is the total useful traffic generated by all content items in a bundle.

\subsection{Traffic gains from bundling}
%Next, we extend the intuition obtained above, and empirically test whether savings can be obtained for bundles of different sizes in our trace. 

  \begin{figure*}
     \centering
			\subfloat[traffic gains with client-side caches \label{fig:caching_results}]
        {\includegraphics[width=0.31\textwidth]{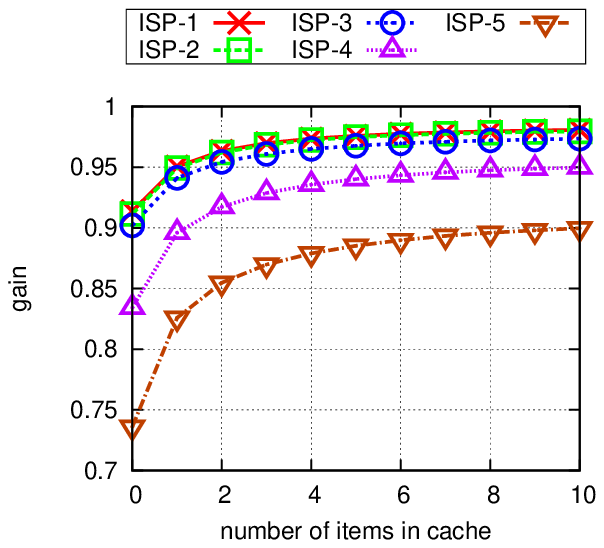}}\hspace{1em}
			\subfloat[factor by which capacity of various content swarms increases with caching\label{fig:caching_coupling}]
        {\includegraphics[width=0.31\textwidth]{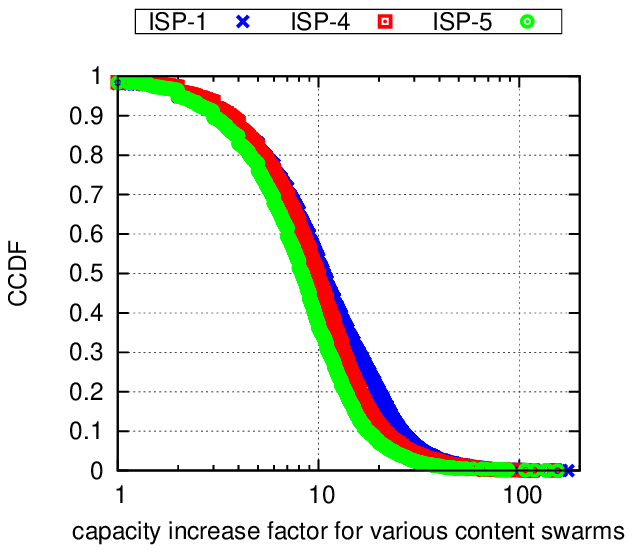}}\hspace{1em}
			\subfloat[traffic gains with client-side caches and constraints on the minimum size of swarms \label{fig:caching_limited}]
        {\includegraphics[width=0.31\textwidth]{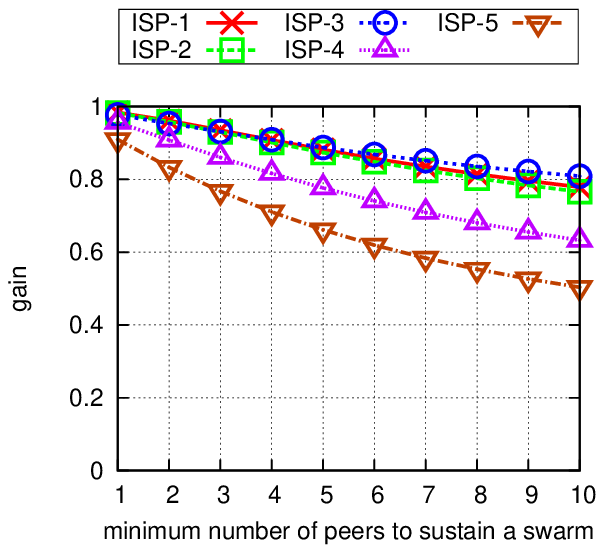}}\hspace{1em}
%			\subfloat[frequency of server accesses for content items with different number of views in different settings\label{fig:caching_unavailiability}]
%     {\includegraphics[width=0.23\textwidth]{figures/availability_popularity_caching.eps}}
%		: without peer-assistence (cdn); with peer-assistance (cdn+p2p); and with historic caches (cdn+p2p+cache) \label{fig:caching_unavailiability}
%    \caption{Characterizing frequency of server accesses for content items with different number of views in different settings: without peer-assistence (cdn); with peer-assistance (cdn+p2p); and with historic caches (cdn+p2p+cache)}
%		\label{fig:caching_unavailiability}
		\caption{Traffic gains in peer-assisted CDNs with historic caches}\vspace{-4mm}
  \end{figure*}

It is worth noting that content bundling has negative impact on traffic savings when the server traffic of a bundle $T^b_s$ exceeds the total server traffic from individual content items $\sum_{i = 1}^k{T^i_s}$. In Figure~\ref{fig:bundling_share}, we consider all possible combinations of items of a given size and estimate a share of those with positive $\Delta G$. Only a minor share of item combinations, i.e., $5-15$\% for combinations of two content items and $9-26$\% for combinations of seven content items, lead to bundles with positive traffic gains. The choice of content items for bundling is even more complicated by the fact that the arrival rates $r_i$ of content items are not known before hand and, so, it is not possible to estimate the traffic savings $\Delta G$ of the bundles at the time the bundles are being formed. Figure~\ref{fig:bundling_gain} shows the average delta gains obtained for the cases when positive savings are obtained. The difference in gain from bundling does not exceed more than $5-7$\% even for very large bundles. 

In summary, bundling appears to not add much additional value, given the already large availability ensured by the ``online while you watch'' model. Bundling also creates additional feasibility concerns in a streaming-like setting.  Among others, bundling assumes altruistic users willing to contribute their traffic and local storage for sharing content items which they may never watch. Similarly, we do not take into account bandwidths required to download and share large content bundles. Therefore, even the asymptotic delta gains of up to $7$\% achieved with our model for content bundling can be affected by other factors. 

%%%%%%%%%%%%%%%%%%%%%%%%%%%%%%%%%%%%%%%%%%%%%%%%%%%%%%%%%%%%%%%%%%%%%%%%%%%%%%%%%%%%%%%%%%%%%%%%%%%%%%%%%%%%%%%%%%%%%%%%%%%%%%%%%%%
%%%%%%%%%%%%%%%%%%%%%%%%%%%%%%%%%%%%%%%%%%%%%%%%%%%%%%%%%%%%%%%%%%%%%%%%%%%%%%%%%%%%%%%%%%%%%%%%%%%%%%%%%%%%%%%%%%%%%%%%%%%%%%%%%%%
%%%%%%%%%%%%%%%%%%%%%%%%%%%%%%%%%%%%%%%%%%%%%%%%%%%%%%%%%%%%%%%%%%%%%%%%%%%%%%%%%%%%%%%%%%%%%%%%%%%%%%%%%%%%%%%%%%%%%%%%%%%%%%%%%%%
\section{Historic Caches}

Historic caching is another common mechanism to improve content availability. Unlike content-bundling, caching is a \emph{reactive} approach which doesn't require any extra traffic from the server as additional availability is induced by the content which has been already seeded to peers. As a result we observe that historic caches are extremely effective in this context. 

\subsection{Traffic gains from caching}

To determine the impact of historic caches on traffic gains we ran a modified version of simulator, in which $k$ content items recently watched by users are cached on their devices and become available to fellow peers next time users come online. In Figure~\ref{fig:caching_results} we plot the traffic gains achieved with these settings as a function of $k$. The traffic gain increases by  $3-12$\% when a single content item (the last item watched) is cached by all users, and gradually grows as the number of last-viewed items cached increases. We note that the results of experiments with unlimited caches show insignificant improvements, i.e., less than $2$\% across all ISPs, with respect to the traffic gains achieved with the size of cache limited to the $k = 10$ last-watched shows. This can be explained by the fact that majority of the BBC iPlayer subscribers are occasional users, i.e., $80$\% of users watch not more than $10$ episodes, and the average number of items in the unlimited cache setting does not exceed $k = 5$. From Figure~\ref{fig:caching_coupling} we note that on average the capacity of swarms with caching is increased by a factor of $\times10$ with respect to the benchmark results when no caching is used. More interestingly, the capacity increases even for the swarms of unpopular content, such as the one we consider in the bottom right plot of Figure~\ref{fig:gain_individual_swarms}.

%Unlike content bundling, client-side caches achieve additional content availability without additional traffic overhead that makes their deployment more feasible.  

	%	\begin{figure}
  %   \includegraphics[width=3in]{figures/isps_limited_caching.eps}
  % \caption{Results of the simulations with unlimited caching and limited upload bandwidth}
  % \label{cd_gain_caching_limited}
 %\end{figure}

%%%%%%%%%%%%%%%%%%%%%%%%%%%%%%%%%%%%%%%%%%%%%%%%%%%%%%%%%%%%%%%%%%%%%%%%%%%%%%%%%%%%%%%%%%%%%%%%%%%%%%%%%%%%%%%%%%%%%%%%%%%%%%%%%%%
%%%%%%%%%%%%%%%%%%%%%%%%%%%%%%%%%%%%%%%%%%%%%%%%%%%%%%%%%%%%%%%%%%%%%%%%%%%%%%%%%%%%%%%%%%%%%%%%%%%%%%%%%%%%%%%%%%%%%%%%%%%%%%%%%%%
%%%%%%%%%%%%%%%%%%%%%%%%%%%%%%%%%%%%%%%%%%%%%%%%%%%%%%%%%%%%%%%%%%%%%%%%%%%%%%%%%%%%%%%%%%%%%%%%%%%%%%%%%%%%%%%%%%%%%%%%%%%%%%%%%%%
\subsection{Contributions of heavy users}

%	\begin{figure}
%	 \begin{center} 
%		 \includegraphics[width=0.31\textwidth]{figures/views_per_user.eps}}
%     \includegraphics[width=0.70\columnwidth]{figures/coupling_ccdf.eps}
%    \caption{Factor by which capacity of various content swarms is increased with caching}
%		\label{fig:coupling}
%	 \end{center}
%  \end{figure}

%Next, we check if contributions of "heavy" users with large caches create a bandwidth bottleneck in the system when additional capacity of a swarm is concentrated around a small number of uploaders with limited upload bandwidths. To this end, we model the behavior of the system when the minimum number $m$ of seeders to sustain a content swarm is greater than one, so that each downloader can benefit from at least $m$ uploaders. From Figure~\ref{fig:caching_limited} we note that caching is extremely effective even when the minimum number of seeders is high, featuring an average increase of $+0.2$ to traffic gains in comparison to the corresponding results without caching (i.e., Figure~\ref{fig:gain_limited}). Thus, the traffic gain for the top five ISPs goes over the mark of $50$\% even for $m = 10$ and reaches $78$\% for the largest ISP. We also note that the delta increase in the gain hardly depends on $m$, suggesting, that the additional capacity is not concentrated around "heavy" uploaders but rather distributed among many peers.   

To determine if the contributions of "heavy" users with large caches create a bandwidth bottleneck in the system when the extra content availability induced by caching is concentrated around a small number of peers with limited upload bandwidth, we ran experiments in which we impose constraints on the number of peers required to support a content swarm (i.e., parameter $m$ in the model). 

The results of the corresponding simulations are presented in Figure~\ref{fig:caching_limited}. We note that caching is extremely effective even for the case when strict constraints are imposed on the minimum number of users needed to sustain a swarm. In comparison  to the corresponding results without caching (i.e., Figure~\ref{fig:gain_limited}), we see an increase of up to $20$\%. As a result, the traffic gain for the top five ISPs goes over the mark of $50$\% even for $m = 10$ and reaches $78$\% for the largest ISP. 

\section{Related papers}
This paper contributes to a line of work~\cite{huang2007can,huang2008understanding,zhao2013peer,balachandran2013analyzing} which has been investigating the feasibility of peer-assisted or hybrid CDNs. We add to this literature by addressing the feasibility of peer assisted streaming of long-duration content, one of the most important applications on the Internet today. Traces from one of the largest deployments of on-demand TV streaming provide a unique opportunity, showing that \emph{at large scale}, a simple ``online while you watch'' assumption can dramatically improve the efficacy of peer assistance, even when various obstacle factors are considered. 

Research into P2P protocols, especially BitTorrent, has long considered obstacle factors which can degrade swarming performance, including ISP-friendliness and locality~\cite{karagiannis2005should,le2011pushing,cuevas2013bittorrent}, and partial participation, which is similar to the concept of free riding~\cite{adar2000free,jun2005incentives,sirivianos2007free}. We borrow from this literature\footnote{It is worth noting that our concept of bitrate stratification is  different from the concept of bandwidth stratification as discussed in BitTorrent literature~\cite{cuevas2013bittorrent}: the former assigns peers to different swarms based on current bitrate encoding, whereas the latter arises as a result of BitTorrent unchoking mechanism, which causes peers of similar download bandwidths to cluster~\cite{legout2007clustering}, even if they are in different ISPs.}, adapting it to peer-assisted CDN swarms, and studying the effect of scale and the ``online while you watch'' model in comparison with the traditional ``online while downloading'' assumption.  In addition, our comprehensive trace also allows us to examine how the characteristics of workload affect different obstacle factors. 

Various analytical approaches were designed to model content availability and peer-matching strategies as ways to increase efficiency of P2P swarming. Among others, Liu \textsl{et al.}\ proposed a model for optimal scheduling in a peer-assisted distribution of user-generated content~\cite{liu2012peer}, Lev~\textsl{et al.}~\cite{lev2010dynamic} analyzed optimal peering strategies using a game-theoretical framework, and Menasche \textsl{et al.}~\cite{menasche2013content} used infinite service queues to model content availability in peer-to-peer swarms. Increasing content availability with bundling was discussed in \cite{menasche2013content, kaune2010seeder, zhang2012dynamic, carlsson2010using, carlsson2012tradeoffs, han2012bundling}. Particularly, Menasche \textsl{et al.}~\cite{menasche2013content} showed that the availability of content bundles decreases exponentially with the size of a bundle, whereas \textsl{Han et al.}~\cite{han2012bundling} reported that bundled content is generally more available among BitTorrent seeders. We build on this work, extending the work of Menasche \textsl{et al.}~\cite{menasche2013content}, and develop a model adapted for peer-assisted CDNs. However, where Menasche \textsl{et al.} focus on content availability,  our focus is instead on gains or savings in server traffic. We also obtain an expression relating gains to the swarm capacity, which yields simple but important insights. 

\vspace{4mm}
\section{Conclusion}

In this paper we studied traffic gains from peer-assisted streaming of long duration content. We  developed a simple analytical model relating the capacity of content swarms to traffic gains, both for individual swarms and aggregated gains in a multi-swarm system. Further, we empirically examined the traffic gains from peer-assisted delivery of on-demand video content using a month-long trace of accesses to BBC iPlayer in London, comprising nearly 16 million sessions and over 2 million users. We studied behavior of the system in the presence of various design obstacles, i.e., isp-friendliness---when content sharing is localized within ISPs, partial participation---when only a part of users opt to re-distribute the content, and bitrate stratification---when there is a need of matching peers with similar bandwidths, and revealed that up to $88$\% of traffic can be saved even despite the obstacles. Our findings also suggest that when operating at scale, a simple ``online while you watch'' model, when users stay online as long as they are watching the content in a manner similar to the user experience in today's CDN/server-based streaming, can be sufficient to secure high gains for peer-assisted CDNs,  provided users contribute/upload to the swarm as long as they are watching the content. 
  %is critical in the presence of high bandwidth peers who can quickly download an item and depart from the swarm. 

%In this paper we have developed an analytical approach to tackle performance analysis of peer-assisted content delivery networks. We have concentrated our attention around the traffic gain metric which aims to assess the amount of traffic offloaded to peers if peer-assistance is deployed. We further validated our model over the realistic traces collected from the BBC iPlayer. Particularly, we have characterized activities of users in the dataset and showed that the gain in the system largely depends on the distribution of content popularity and length. The results of modeling confirmed the previous findings suggesting that the remarkable percentage of traffic, e.g., up to $90$\% can be potentially saved even if the traffic is localized inside large ISPs. 

We also investigated the impact of two well-known techniques for improving content availability on the traffic gains and showed that bundling is not effective in this context as the traffic overhead from downloading large bundles surpasses the benefits of improved availability. In contrast, we observed that a simple caching approach can boost the gain from peer-assistance for up to $23$\%.

Our study focused on on-demand streaming of  \emph{long duration} content (TV shows), which are increasingly dominant in current Internet. Thus, these results confirm the enormous potential of peer-assistance in future content delivery scenarios.

\bibliographystyle{abbrv}
%\vspace{-4mm}
%{\scriptsize;
%\bibliography{biblio}}

\vspace{4mm}
\bibliography{biblio}
\end{document}